\documentclass[final,5p,times,twocolumn]{elsarticle}

\usepackage{graphicx}
\usepackage{amssymb}
\usepackage{mathrsfs}

\usepackage{lineno}
\usepackage{listings}

\usepackage{amsmath,amsfonts,amsthm} 
\usepackage[dvipsnames]{xcolor}
\usepackage{enumitem}
\usepackage{booktabs}

\usepackage{threeparttable}
\usepackage{hyperref}

\usepackage{soul}
\usepackage{ulem}

\definecolor{mygreen}{rgb}{0,0.6,0}
\definecolor{mygray}{rgb}{0.5,0.5,0.5}
\definecolor{mymauve}{rgb}{0.58,0,0.82}
\biboptions{square}

\bibpunct{(}{)}{;}{a}{}{,} 

\journal{arXiv}

\begin{document}

\begin{frontmatter}


\title{\texttt{SCORPIO}, a package for the visualization of galaxy pairs}




\author[iate,oac,famaf]{Jose A. Benavides}
\author[iate,oac]{Martín Chalela}
\author[iate,conae]{Juan B. Cabral}
\author[duke]{Bruno O. S\'anchez}
\author[iate,oac,wsu]{Sebastian Gurovich}

\address[iate]{
   Instituto de Astronom\'ia Te\'orica y Experimental -
   Observatorio Astron\'omico de C\'ordoba (IATE, UNC--CONICET),
   C\'ordoba, Argentina.}
\address[oac]{
    Observatorio Astron\'{o}mico de C\'{o}rdoba, Universidad Nacional de C\'{o}rdoba, Laprida 854, X5000BGR, C\'{o}rdoba, Argentina
}
\address[famaf]{
	Facultad de Matem\'atica, Astronom\'{\i}a y F\'{\i}sica,
    Universidad Nacional de C\'ordoba (FaMAF--UNC)
	Bvd. Medina Allende s/n, Ciudad Universitaria,
    X5000HUA, C\'ordoba, Argentina 
}
\address[conae]{
Comisi\'on Nacional de Actividades Espaciales (CONAE), 
Ruta Provincial C45 a 8 Km, Falda del Ca\~nete,
C\'ordoba, Argentina.
}
\address[duke]{
    Department of Physics, 
    Duke University, 
    120 Science Drive, 
    Durham, NC, 27708, USA
}
\address[wsu]{
    Western Sydney University, Locked Bag 1797, Penrith South DC, NSW 2751, Australia
}

\begin{abstract}

We present the description of the project \texttt{SCORPIO}, a Python package for retrieving images and associated data of galaxy pairs based on their position, facilitating visual analysis and data collation of multiple archetypal systems.
The code ingests information from SDSS, 2MASS and WISE surveys based on the available bands and is designed for studies of galaxy pairs as natural laboratories of multiple astrophysical phenomena such as tidal force deformation of galaxies, pressure gradient induced star formation regions, morphological transformation, to name a few.

\end{abstract}

\begin{keyword}
galaxies: interactions; surveys; techniques: image processing; python packages


\end{keyword}
\end{frontmatter}



\section{Introduction}

The cosmological model of cold dark matter with a cosmological constant establishes a context in which galaxies and galaxy systems form hierarchically through merger events. In this scenario interacting galaxies are excellent laboratories that provide information on the general evolution of galaxy systems in their environment. However sometimes determinations of the level of interaction can be biased by projection effects \citep{Wang_2019}. Here multi-band images both in the optical and infrared part of the EM spectrum of these objects allow characterization of the interaction parameters as well as their star formation history.

Galaxy pairs represent the majority of interacting systems. This can be seen in the decreasing character of the member multiplicity function of groups \citep{gott1977groups}. As they are so abundant and occur mainly in low-density environments, interaction is not much affected by external agents (as could occur within a group or a galaxy cluster environment). For example, in the case of disk galaxies, some characteristics reveal the intensity of interaction like tidal tails \citep{Toomre_and_Toomre_1972}, gas loss due to ram pressure \citep{Gunn_Gott_1972} and other deformations caused by gravitational action. Using spectroscopic data from the Sloan Digital Sky Survey Data Release 4 (SDSS DR4), \cite{ellison2010} analyzed the effects of the environment on the interaction of nearby pairs of galaxies ($r_{\rm{p}} ~ < ~ 80  ~ h^{- 1} kpc, \Delta v ~ < ~ 500 ~ km ~ s^{ - 1}$) and the possibilities of merge events between them who show that in low-density environments the shorter projected distances and lower velocity desperation systems represent a higher fraction of interacting galaxies in contrast to the high-density where these encounters are scarce.

The classification of the type of interaction in galaxy pairs can be addressed by two distinct strategies, automatic methods and human visual inspection. An example of the later case is the Galaxy Zoo Project \citep{keel2013}, where the help of the general public is used to determine morphological and interacting aspects. The use of automatic methods, i.e. machine learning techniques, is increasingly relevant to the sheer amount of data produced by large mega-surveys as well as improvement in numerical simulations, providing a means for statistical analyses of very large homogeneous data samples \citep[e.g.][]{Shamir2014, Ferreira2020, Prakash2020}.

Different authors have used data from different surveys to identify pairs of interacting galaxies in different environments \citep{Alonso_2004, Alonso_2012, Duplancic_2018, Patton_2016, Rodriguez2020}. In this context, SCORPIO\footnote{\url{https://github.com/josegit88/SCORPIO/}} (Sky COllector of galaxy Pairs and Image Output) is a Python open-source project that aims to provide a tool for fast visual inspection of galaxy pairs in multiple surveys.
The work is structured as follows: in Sec.~\ref{meteoric_details} some theoretical details and considerations for the generation of  information of pairs of galaxies are given. In Sec.~\ref{app_scorpio} the main functions and characteristics of \textsc{SCORPIO} are described that allows for the generation of the desired image for visualization. Some examples of the application of the code to reported pairs of galaxies are shown in Sec.~\ref{scorpio_images} and Sec.~\ref{application}. Finally, in Sec.~\ref{conclusiones} some conclusions and perspectives of the project are presented.

\section{Methodology}\label{meteoric_details}

A large number of properties of different objects are analyzed from information obtained from sources in different filters. One technique involves stacking images of the same object at various wavelengths. This may involve smoothing all images to a common resolution that is useful to reveal both information of individual objects as well as to their environment, as applied in \citet{zibetti2005} who find a separation between BCG and ICL properties. Likewise, sometimes this exercise of stacking information can lead to confusion, which must be detected and prevented using synthetic mocks that allow cleaning the data for possible bias \citet{jones2016}. In this sense, it is important to have some complementary information of the environment when studying many astrophysical systems that are diffuse and spatially extended. In our case for the interaction of galaxy pairs, candidates may be visually inspected to present characteristic interaction effects and to mitigate false-positive pairs caused mostly by projection effects. This can be limited, for example if there is spectroscopic information for each object which allows distance determination as well as traces of a possible previous interaction of regions where gas is compressed, heats up, giving rise to these bursts of star-formation, observable in $H_{\alpha}$, that con be modelled using spectro-photo-metric stellar population models and for certain narrow-band and broad-band filters such as IR \citep{keel1985, dopita2002}. Some affects of the interaction physics on star formation has also been studied using numerical simulations \citep{perez2006}.

\subsection{Data retrieval}

For the stacking process, photometric images are retrieved from public databases using specific software designed for querying astronomical images. In particular, we use the Astroquery \footnote{\url{https://astroquery.readthedocs.io/en/latest}} service which provides a querying interface with several databases using a unified API (Application Programming Interface) when possible. With SCORPIO we provide ease of interaction with the following three public sky surveys of wide-spread use:
\begin{itemize}
    \item {\bf Sloan Digital Sky Survey (SDSS)}: Provides the largest photometric database available at present in the visible spectrum, covering 14055 square degrees of sky imaged in five bands ($u$, $g$, $r$, $i$ and $z$) with a limiting magnitude $r=22.2$ \citep{SDSS-DR16}.
    \item {\bf Wide Field Infrared Survey Explorer (WISE)}: Provides an all-sky infrared wavelength mapping in four bandpasses: 3.4, 4.6, 12, and 22 $\mu m$, with angular resolutions of $6.1"$, $6.4"$, $6.5"$ and $12.0"$ respectively \citep{wise}.
    \item {\bf Two Micron All Sky Survey (2MASS)}: Provides an all-sky near-infrared mapping in three bandpasses: $J \ (1.25 \mu m)$, $H \ (1.65 \mu m)$ and $K_s \ (2.16 \mu m)$ \citep{2MASS}.
\end{itemize}

Once the photometric data of the sky region of interest is obtained, the final images centered on the science objects or galaxy pair are generated automatically for the user. In the case that multiple filters are requested, the stacked image for every available filter is computed in order to present a single composite image.


\subsection{Pair separation}

The co-moving distance from the observer to an object in The Universe is given as a function of the redshift (without considering peculiar velocities) and the cosmological parameters by the expression \citep{Mo}:

\begin{equation}
    d(z) = \frac{c z}{H_0} [ 1 + \mathcal{F}_d (z ; \Omega_{m,0}, \Omega_{\Lambda , 0}, ...) ]
	\label{eq:dist_z}
\end{equation}

where $\mathcal{F}_d$ is a function that depends on the redshift such that $\mathcal {F}_d \ll 1 $ for $z \ll 1$, in which case we obtain the classic Hubble law $cz = H_0 d$. For brevity only $\Omega_{m, 0}$ (matter density) and $\Omega_{\Lambda, 0}$ (dark energy density) are written but $\mathcal{F}_d$ depends on the full set of cosmological parameters. As will be detailed in Section \label{app_scorpio}, the user can specify a desired cosmology using the Astropy \citep{astropy_a, astropy_b} package.

With the comoving distance value (between the observer and the pair of galaxies) and the angular separation between galaxies ($\theta$, computed from the celestial coordinates right ascension and declination) on the sky, the physical separation between them is given by:

\begin{equation}
    r = a \, d(z) \, \theta
	\label{eq:separation}
\end{equation}
Where $a$ is the scale factor and is given by the redshift as $a = (1 + z)^{-1}$.

Distance information is then projected onto the final image to help the user rapidly characterize the physical interaction parameters between the galaxies and between the galaxies and environment.

\section{Structure of SCORPIO}\label{app_scorpio}

\subsection{General description}

SCORPIO offers a mediator service between the user and databases with photometric image information, receiving the position coordinates (along with other optional parameters) and quickly returning an image (with  customizable format, size and visualization color).

The package was developed for Python 3.7 or higher, and is built on-top of the following Python scientific-stack libraries: Astropy \citep{astropy_a, astropy_b} for coordinates and cosmological computations, Astroquery \citep{astroquery} for querying astronomical databases, Matplotlib \citep{matplotlib} and Seaborn \citep{Waskom2021seaborn} for high-level image creation, NumPy \citep{numpy} for array manipulations and Attrs \footnote{\url{https://www.attrs.org}} to reduce boilerplate code in the implementation of classes.

\subsection{Technical details}

\texttt{SCORPIO} API provides several functions that allows the user to compute different quantities of interest that serve to generate a final image with relevant descriptive information.

The main functionality can be summarized in the following methods:
\begin{itemize}
    \item \texttt{gpair}: This function wraps the entire functionality of \texttt{SCORPIO}, receiving galaxy pair coordinates as input and returning an image object (an instance of the class GPInteraction). The user can specify the sky survey and photometric filters from which to retrieve the information. In the case where multiple filters are requested, a stacked composite image is generated.
    
    \item \texttt{GPInteraction}: This object stores the desired information of the galaxy pair as attributes (such as the image array or the header) and provides the method \texttt{.plot()} that generates a figure with a summary of the relevant information.
\end{itemize}

Given that the main function \texttt{gpair} performs many tasks, these are built as individual functions. We implement them as part of the public API of \texttt{SCORPIO} with their corresponding documentation, to provide the user with enough flexibility to execute each task individually. These functions are the following:

\begin{itemize}
    \item \texttt{download\_data}: handles the query and download steps given the specified survey using the Astroquery library.
    
    \item \texttt{stack\_pair}: combines the complete set of images from both galaxies and their specified filters into a single image.

    \item \texttt{distances}: computes the physical and pixel distances between both galaxies. The user can provide a custom cosmology as defined by the \texttt{astropy.cosmology} module. If no cosmology is provided as an input, \texttt{SCORPIO} will use the astropy default cosmology (\texttt{Planck18} since version 4.2).
\end{itemize}

A detailed description of the input parameters and functionality of each method is provided in the docstrings.

\subsection{Quality assurance}

To provide a reliable package we followed modern software engineering practices throughout the entire development process. We adopted the \textit{PEP 8 -- Style Guide for Python Code} \citep{PEP8} to ensure readability and consistency. Also, the tools flake8\footnote{ \href{https://flake8.pycqa.org/en/latest/}{https://flake8.pycqa.org/en/latest/}} and black\footnote{ \href{https://pypi.org/project/black/}{https://pypi.org/project/black/} } were used for code linting to ensure there are no deviations from the PEP 8 norm.

We developed a comprehensive unit-test suite \citep{Jazayeri2007} to ensure that \texttt{SCORPIO} works as expected and that fundamental errors have been handled, achieving a code-coverage \citep{MillerMaloney1963} of 97\%. The testing suite is tested for Python versions 3.7, 3.8 and 3.9. An online documentation is automatically compiled from code docstrings and hosted by the read-the-docs\footnote{ \href{https://scorpio-rdd.readthedocs.io/en/latest/}{https://scorpio-rdd.readthedocs.io/en/latest/} } service. A tutorial explaining the more advanced features and configurations is also included in the online documentation.

Finally, SCORPIO source code is under the MIT License \citep{MITLicense} and publicly available in its GitHub repository\footnote{ \href{https://github.com/josegit88/SCORPIO}{https://github.com/josegit88/SCORPIO} }. The entire development process is supervised by GitHub's continuous-integration service\footnote{ \href{https://github.com/josegit88/SCORPIO/actions/workflows/scorpio_ci.yml}{https://github.com/josegit88/SCORPIO/actions/workflows/scorpio\_ci.yml} } which permanently checks the build status of the project;
and also the project is currently going through registration/evualuation process of the the Astrophysics Source Code Library (ASCL.net, \citealp{grosbol2010making}) 

\begin{figure}
\includegraphics[width=\columnwidth]{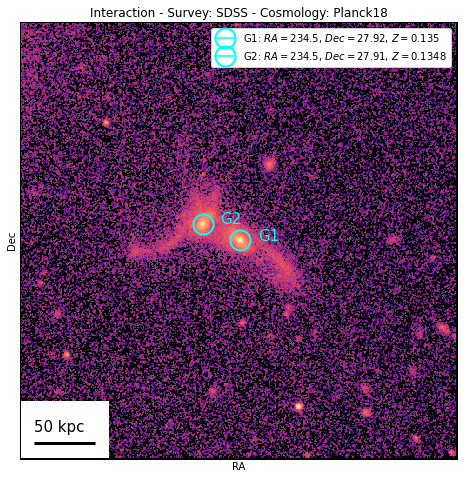}
\caption{Example of the image generated for an interacting galaxy pair, using filters g,i from SDSS. The system is located at a redshift $\sim 0.135$. The image is centered at the primary galaxy as indicated with the first galaxy introduced in the function \texttt{gpair}. The color scale is normalized to the pixel count of the primary galaxy.}
\label{fig:example_image}
\end{figure}

\subsection{View example}\label{scorpio_images}

As a case study for \texttt{SCORPIO} we use a simple example of a galaxy pair selected from \citet{Rodriguez2020}, here we choose a pair in which the tidal tails show signs of previous interaction. The following script shows how to use \texttt{SCORPIO} to generate an image of the selected galaxy pair from its member coordinates (right ascension, declination and redshift). The resulting image is shown in Figure \ref{fig:example_image}. A detailed description of how to customize the resulting image can be found in the online documentation as well as an advanced section to generate a publication-quality figure.

\begin{lstlisting}[language=Python]

>>> # import scorpio module
>>> import scorpio

>>> # generate the image object
>>> gpi = scorpio.gpair(
...    # first galaxy
...    ra1=234.48,
...    dec1=27.92,
...    z1=0.1349,

...    # second galaxy
...    ra2=234.48,
...    dec2=27.91,
...    z2=0.1348,

...    # survey and image resolution
...    survey="SDSS",
...    resolution=500,
... )

>>> # plot with default configuration
>>> gpi.plot()
\end{lstlisting}

\section{Application to a catalog of galaxy pairs}\label{application}

\begin{figure}
\includegraphics[width=\columnwidth]{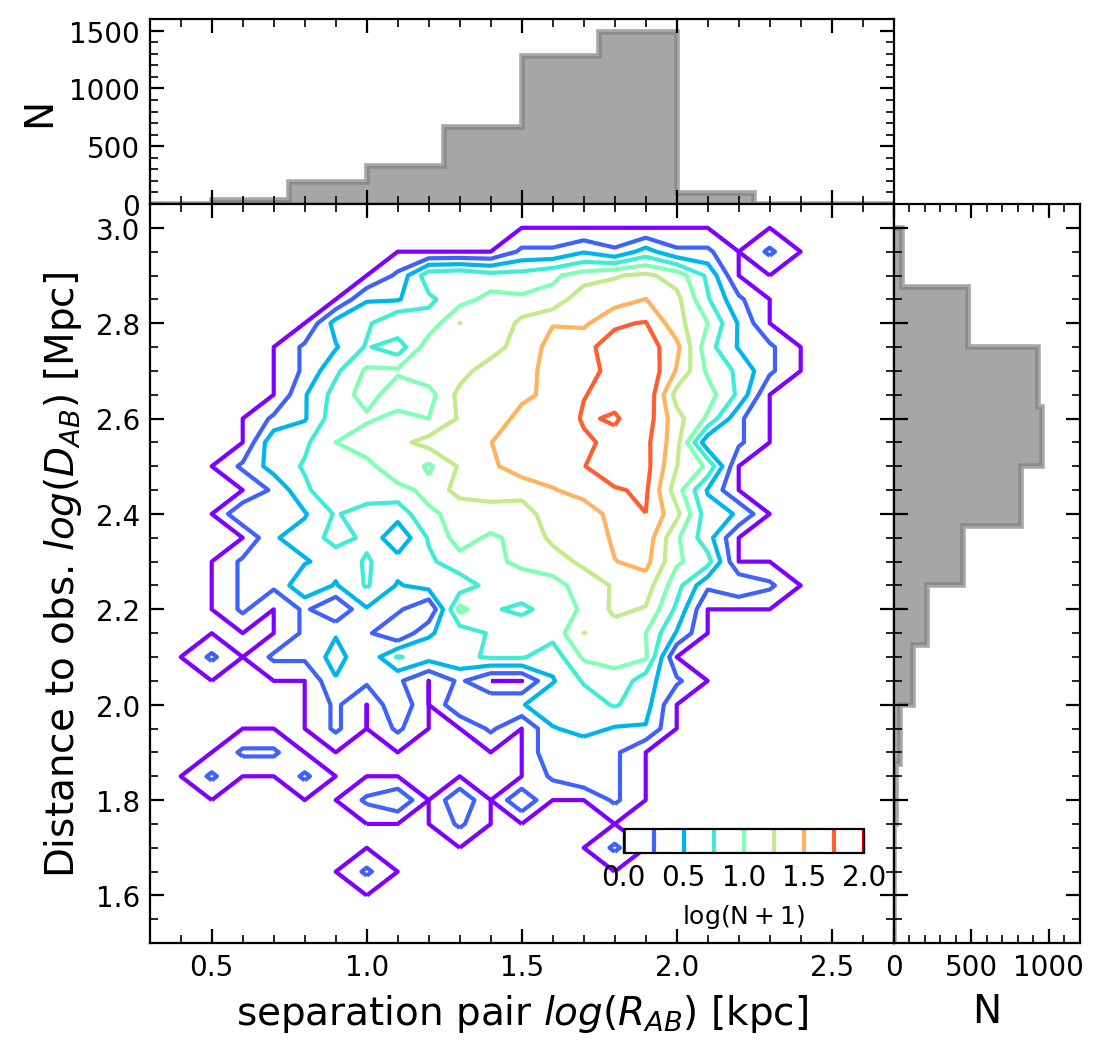}
\caption{Relation between distance to the observer $(D_{AB})$ and mean separation of the galaxy pairs $(S_{AB})$, for the sample of 8182 pairs of galaxies, from the \citet{Rodriguez2020} catalog.}
\label{fig:distances}
\end{figure}

As a practical application example, we use the data from the \citet{Rodriguez2020} catalog which contains a sample of more than 8000 pairs of nearby galaxies (whose redshift measurements range between $ z = $ 0.0105 and $ z = $ 1.9967 ) from data from the Sloan Digital Sky Survey (SDSS, \citet{SDSS}) and taking into account features such as velocity dispersions, distance and apparent magnitude between of galaxy pair, also, they were taken into account the isolation criteria of this system with other objects.
A relationship between the relative distances between the pair and the distance to the observer is presented in Fig.~\ref{fig:distances}, where we can see that most of these pairs are located at distances estimated at approximately between $250$ and $600$ Mpc, with mean separations between $31$ and $63$ kpc (both estimates being very sensitive to redshift).

\begin{figure*}
\centering
\includegraphics[width=\textwidth]{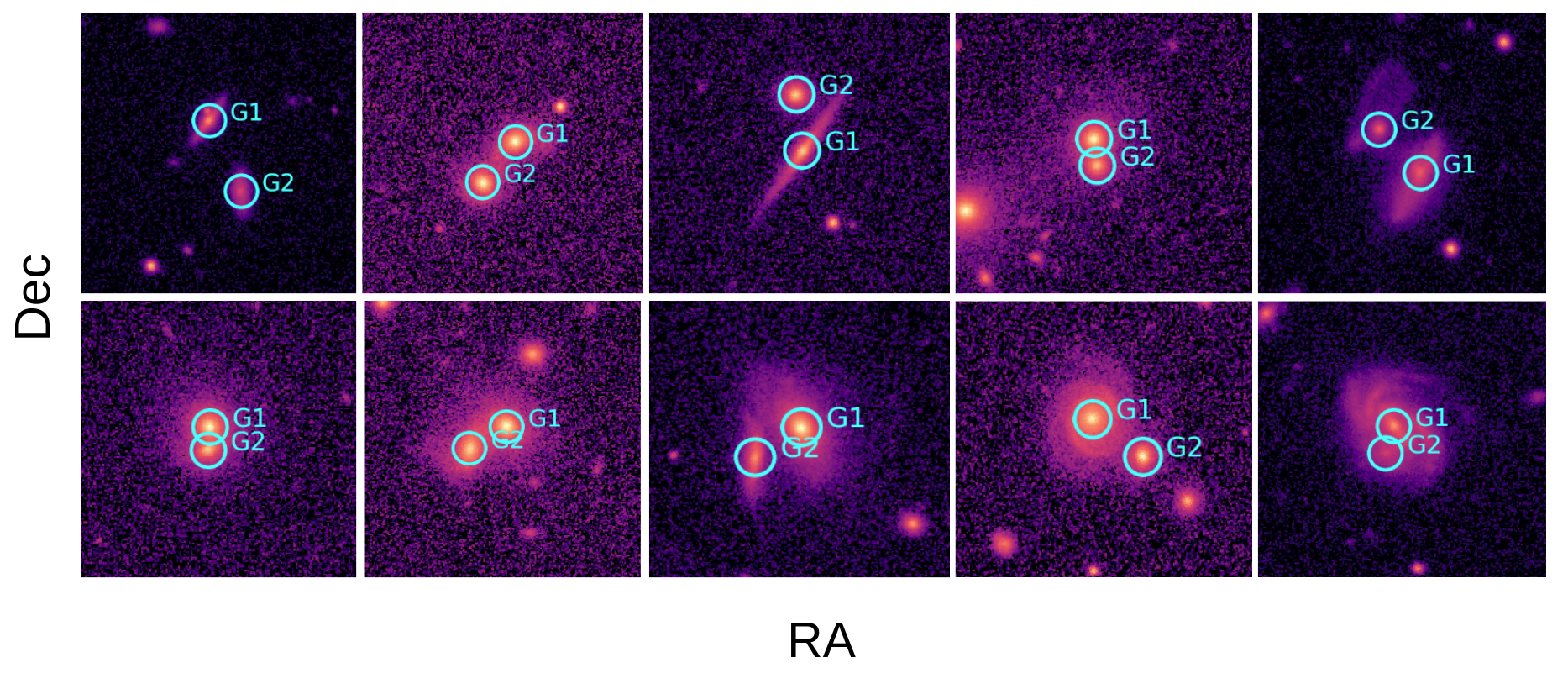}
\caption{Example of the view generated with \texttt{SCORPIO}, for ten random cases of the 400 nearby galaxy pair sub-sample, with separation between 10 and 30 kpc and distances between 300 and 500 Mpc. Tidal tails are clearly evident for some pairs  indicating these pair of interacting galaxy candidates are genuine  and not unrelated pairs aligned only due to projection effects. }
\label{fig:sample_view}
\end{figure*}

Interacting galaxy pairs are excellent test laboratories relevant to many areas of astrophysics. In order to demonstrate batch use of Scorpio we make a small selection of close pairs, with separation $\log(S_{AB}/kpc) = [1,1.5]$ and distances $\log(D_{AB} / Mpc) = [2.5, 2.7]$, corresponding to a subsample of $\sim$ 400 pairs. Ten of these pairs where chosen at random and their images generated with \texttt{SCORPIO}, in the filters (g, r) are presented in Fig.~\ref{fig:sample_view}. It can be seen visually, how some pairs show characteristics of dynamical interactions such as tidal tails. 

\section{Conclusions}\label{conclusiones}

We present the first version of SCORPIO, an open source package developed in Python that aims to facilitate the visual analysis of astronomical images of galaxy pairs. The development process of SCORPIO was carried out using modern software engineering tools and techniques to ensure that we are providing a robust package for the astronomical community to refine and update.

The main aspects of SCORPIO can be summarized in the following conclusions:
\begin{itemize}
    \item SCORPIO performs queries to multiple sky surveys to retrieve and combine photometric images of galaxy pairs using a very simple and intuitive user interface. For the moment, SCORPIO can perform queries to SDSS, WISE and 2MASS databases.
    
    \item The user can quickly generate publication-quality plots of the desired galaxy pair with relevant information, such as physical separations. This can be a powerful solution in studies that rely on visual inspections of systems where the type of interaction between galaxies needs to be addressed, e.g. "merging system", "interacting" or "non-interacting".
\end{itemize}

Finally, for future releases of SCORPIO, we will implement more functionalities that aim to provide the user with additional relevant information, such as luminosity or morphological classification that is already present in some databases. We will also explore the possibility of retrieving spectroscopic information for each galaxy when data become available, and also integrate spectrum plotting methods to SCORPIO.

\section{Data availability}
The catalogs of galaxy pairs used to carry out the tests in the development of SCORPIO were requested directly from the authors of the referenced works \citet{Rodriguez2020, Alonso_2012}, these can be shared upon request to the corresponding author if there is no further conflict with the projects in progress.

\section{Acknowledgments}
This work was partially supported by the Consejo Nacional de Investigaciones Cient\'ificas y T\'ecnicas (CONICET, Argentina). JAB was supported by a fellowship from Secretar\'ia de Ciencia y Tecnolog\'ia (SeCyT, Argentina), MC and JBC were supported by a fellowship from CONICET.

The authors would like to thank M.S. Alonso, F. Rodriguez and collaborators for kindly providing the galaxy pair catalogues used in the development of this project.

This research employed the http://adsabs.harvard.edu/, Cornell University arXiv.org repository, the Python programming language, the Numpy and Scipy libraries, and other packages that can be found at the GitHub repository for SCORPIO.


\bibliographystyle{aa}
\bibliography{bibliography.bib}


\appendix


\end{document}